\documentclass[journal]{IEEEtran}
\usepackage{nicematrix}
\usepackage{color}
\usepackage{cite}
\usepackage{textcomp}
\usepackage{xcolor}
\usepackage{balance}
\usepackage{graphicx}
\usepackage[english]{babel}
\usepackage{amsthm,amssymb}
\usepackage{amsmath}
\usepackage{subfigure}
\usepackage{mathtools}
\usepackage{tikz}
\usepackage{pgf}
\usepackage{amsfonts}
\usepackage{bm}
\usepackage[bottom]{footmisc}
\usepackage[ruled,linesnumbered]{algorithm2e}
\usepackage[T1]{fontenc}
\usepackage{bbm}
\usepackage{enumerate}

\usepackage{setspace}
\usepackage{array}
\usepackage{multirow}
\usepackage{diagbox}

\newtheorem{lemma}{Lemma}

\newtheorem{theo}{Theorem}
\newtheorem{dfn}[theo]{Definition}

\allowdisplaybreaks
 
\def\BibTeX{{\rm B\kern-.05em{\sc i\kern-.025em b}\kern-.08em
    T\kern-.1667em\lower.7ex\hbox{E}\kern-.125emX}}
\begin{document}

\title{A Graph-based Strategic Sensor \\ Deployment Approach for $k$-coverage in WSN }

\author{\IEEEauthorblockN{Lakshmikanta Sau, Priyadarshi Mukherjee, and Sasthi C. Ghosh}

\IEEEauthorblockA{Advanced Computing \& Microelectronics Unit\\
Indian Statistical Institute, Kolkata 700108, India\\
Emails: lakshmikanta\_r@isical.ac.in, priyadarshi@ieee.org, sasthi@isical.ac.in}}
\maketitle
\begin{abstract}
    This paper studies a graph-based sensor deployment approach in wireless sensor networks (WSNs). Specifically, in today's world, where sensors are everywhere, detecting various attributes like temperature and movement, their deteriorating lifetime is indeed a very concerning issue. In many scenarios, these sensors are placed in extremely remote areas, where maintenance becomes challenging. As a result, it is not very wise to depend on a single sensor to obtain data from a particular terrain or place. Hence, multiple sensors are deployed in these places, such that no problem arises if one or few of them fail. In this work, this problem of intelligent placement of sensors is modelled from the graph theoretic point of view. We propose a new sensor deployment approach here, which results in lesser sensor density per unit area and less number of sensors as compared to the existing benchmark schemes. Finally, the numerical results also support our claims and provide insights regarding the selection of parameters that enhance the system performance.    
\end{abstract}
\section{Introduction}
With the rapid evolution of applications like Internet of Things (IoT), wireless sensors are everywhere. Wireless sensor network (WSN) is a reality today; the global IoT connections is forecast to reach around $38.9$ billion in $2029$ \cite{ericsson}. As a result, the lifetime of these sensors is a very critical issue and hence, they need to be regularly monitored and recharged. In this context, there has been research regarding efficient environment aware wireless transmission techniques, which extend the sensor lifetime by avoiding unwanted transmissions \cite{cl},\cite{irfan}. However, in many applications, they are placed in very remote and challenging areas, where their maintenance is extremely challenging. Thus, it is not wise to rely solely on a single sensor placed in such areas for data. In graph theoretical terms, this is termed as $1$-coverage, i.e., every point in the region is monitored or covered by at least one sensor.

This leads to a more evolved concept of graph theory, which is $k$-coverage, i.e., all the points in a concerned `region' is covered or monitored by at least $k$ sensors. Note that, the value of $k$ depends on the application at hand; a higher value of $k$ implies that a higher degree of reliability is required, and vice-versa \cite{surv}. The simplest possible strategy to attain a certain coverage is the one, where we deploy all the available sensors in the region of concern, such that the defined set of targets are totally covered. However, as stated earlier, these sensors have a limited lifetime and hence, this method is not desirable from the application point of view \cite{access}. Also, this leads to unwanted wastage of resources. The authors in \cite{k-cov1} proposed a bound on the sensor density in order to cover a certain region of interest. The work in \cite{k-cov} proposed a sensor deployment strategy, where they used regular hexagons of a certain side length to cover the entire region and accordingly, place the sensors inside.

In this context, we propose an intelligent graph-based sensor deployment strategy. Specifically, Section II introduces the essential terminology required throughout the work and also the respective system model.  Section III discusses the proposed strategy by modeling the sensor deployment problem as a $k$-coverage problem, where it is assumed that the sensors are placed in the environment as a regular polygon tiling. To do so, we assess and examine various planar convex regular polygons, which possess the ability to tile the entire Euclidean plane. As hexagons prove to be the most efficient of the regular polygons, we assume hexagonal tiling. Based on hexagonal tiling of the region of interest, we strategically place the sensors inside a regular hexagon, and by employing this method for every hexagon, we cover the entire Euclidean plane. Moreover, we characterize the performance of the proposed strategy in terms of sensor density and number of sensors required to cover a certain region of interest. We show that our proposed strategy outperforms the state of the art deployment strategy \cite{k-cov} in terms of both sensor density and total number of sensors required. Finally, Section IV presents the numerical results and Section V concludes our work.

\section{System Model and Preliminaries}
Here we first introduce the essential terminology, which is used throughout the work and then, the network model.

\subsection{Definitions}
\begin{dfn}(Regular polygon) :
A closed geometric figure formed by edges, not necessarily equal, is called a polygon. Moreover, a polygon with equal edges and identical internal angles is said to be a regular polygon.
\end{dfn}
 \begin{dfn}(Regular polygon tiling) :
The act of tiles covering up an Euclidean plane is called tiling, where each point on the Euclidean plane is covered by at least one tile. If the Euclidean plane is covered by regular polygons, then it is called a regular polygon tiling.
     
 \end{dfn}  
 \begin{dfn}(Cover set) :
 If a set of regular polygons covers an Euclidean plane, then this particular set of polygons is said to be a cover set of this Euclidean plane. Here, an Euclidean plane is said to be covered if each point on the plane is covered by at least one regular polygon. 
 \end{dfn}
 \begin{dfn}($k$-coverage set) :
 A cover set is defined as a $k$-coverage set if each point of an Euclidean plane is covered by at least $k$ number of regular polygons. 
 \end{dfn}
 \begin{dfn}(Sensing range) :
In a WSN, each sensor monitors the physical conditions of the environment and/or the wireless network up to a certain distance, i.e., it can sense its desired attribute within a circle of fixed radius $r$. This circle (with radius $r$) is defined as the sensing range of the sensor.
 \end{dfn}
\begin{dfn}(Communication range) :
A sensor monitors  the physical conditions within its sensing range and communicates the desired attribute to a destination. In this context, the communication range of a sensor $s$ is defined as the maximum distance from $s$, within which it can communicate directly with another sensor.  
\end{dfn}

\begin{dfn}(Adjacent polygons) :
    Two regular hexagons, of equal sides, are said to be adjacent to each other if they share a common edge. 
\end{dfn}

\begin{dfn}(Overlapping polygons) :
    Two regular hexagons, of equal sides, are said to be overlapping, if they contain at least one common point, which lies inside both the hexagons. 
\end{dfn}


\begin{figure}
    \centering
    \includegraphics[width=.72\linewidth]{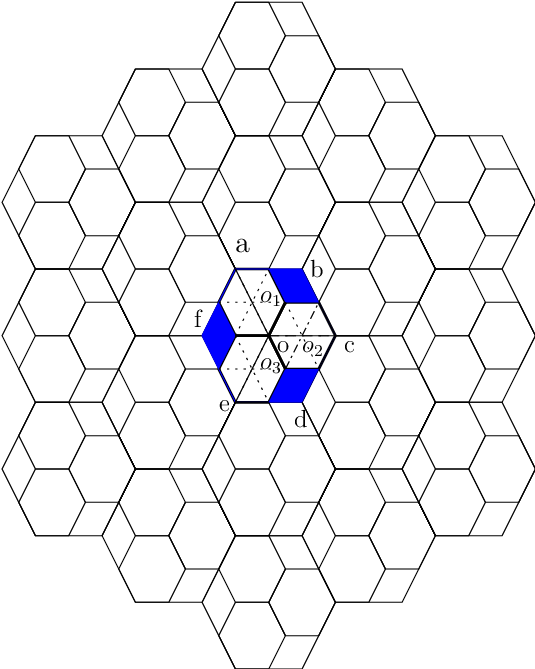}
    \caption{Two different tilling of Euclidean plane.}
    \vspace{-4mm}
    \label{sys}
\end{figure}
\subsection{Network Model}
In a typical WSN, sensors are placed throughout the environment to monitor certain physical conditions. In many cases, they are placed in extremely remote areas to collect useful data and as a result, their maintenance becomes extremely challenging in that scenario. Therefore, it is not wise to depend on a single sensor to obtain the useful data from a particular point. As a result, we want each point of a particular place to be covered by multiple sensors, such that there is no problem in case one or some of them fail. We model this as a $k$-coverage problem, i.e., each point of a particular area in the Euclidean plane is covered by at least $k$ regular polygons. Here, we assume that each sensor of the WSN have similar characteristics and follow identical lifetime models. We further assume, that all the sensors  are aware of their own position with respect to the entire network and also, they possess full knowledge of all the other sensors by means of a common central system. Finally, we characterize the sensing range of sensor $s_i$ as $SR_i$, which is a disk of radius $r$ and the corresponding communication range is modeled as $CR_i$ (a disk of radius $R$).

 
\section{Proposed Strategy}

In this section, we discuss the proposed sensor deployment strategy. As stated earlier, we model this scenario as a $k$-coverage problem, where we assume that the sensors are placed in the environment as a regular polygon tiling.  Note that, this is not an ideal solution if the sensors are randomly deployed and thus, we propose to place the sensors strategically, such that the resulting cover set offers $k$-coverage with fewer sensors. Moreover, as shown in Fig.\ref{sys}, now we examine and assess a collection of planar convex regular polygons, which possess the ability to tile the Euclidean plane.

\begin{figure}[t!]
    \centering
    \includegraphics[width=0.88\linewidth]{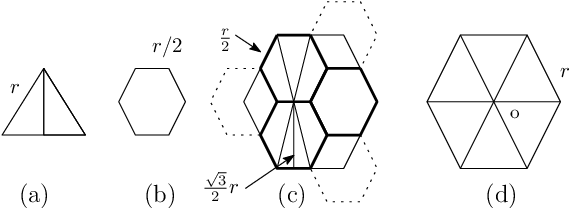}
    \caption{Different shape tiles}
    \vspace{-4mm}
    \label{con_fig}
\end{figure}

Now, examining these planar convex tiles is crucial to understand the $k$-coverage issue in planar WSNs, where Fig. \ref{con_fig} demonstrates a few convex regular polygons. Specifically, Fig. \ref{con_fig}(a) represents an equilateral triangle with side $r$, i.e., the area of this triangle is $\frac{\sqrt{3}}{4} r^2$ and Fig. \ref{con_fig}(b) represents a regular hexagon $H_{r/2}$ of side $r/2$. Since $H_{r/2}$ consists of six equilateral triangles with side $r/2$, its area is obtained as
\vspace{-2mm}
\begin{equation}\label{area_r/2}
    A(H_{r/2})=6\times \frac{\sqrt{3}}{4} (r/2)^2 =\frac{3\sqrt{3}}{8}r^2.
\end{equation}

Fig. \ref{con_fig}(c) represents a regular hexagon $H_{r}$ with side $r$. Later, we demonstrate that $H_{r}$ can contain a maximum of three non-overlapping $H_{r/2}$. Finally, Fig. \ref{con_fig}(d) illustrates a $H_{r}$, which consists of six equilateral triangles of side $r$, i.e., regular polygon tiling of six equilateral triangles. Therefore, the area of $H_{r}$ is 
\vspace{-2mm}
\begin{equation}\label{area}
    A(H_{r})=6\times \frac{\sqrt{3}}{4} (r)^2 =\frac{3\sqrt{3}}{2}r^2
\end{equation}
Though, from \eqref{area_r/2} and \eqref{area}, we observe that $A(H_{r})=4\times A(H_{r/2})$, in the following lemma, we prove that more than three $H_{r/2}$ can not completely lie within a $H_{r}$.
\begin{lemma}\label{lema1}
    A hexagon $H_{r}$ of side $r$ can contain a maximum of three non-overlapping hexagons $H_{r/2}$, each of side $r/2$.
\end{lemma}
\begin{proof}
    The length of the largest diagonal of $H_{r}$ and $H_{r/2}$ is $2r$ and $r$, respectively, i.e., a single $H_{r/2}$ can completely lie within $H_{r}$. Hereafter, let $H^1_{r/2}$ and $H^2_{r/2}$ be two non-overlapping hexagons of side $r/2$ each, and $X=H^1_{r/2} \cup H^2_{r/2}$. Hence, the furthest distance between two points of $X$ is at least 
    \vspace{-2mm}
    \begin{equation}\label{fd}
        l(X)=2\sqrt{(\sqrt{3} (r/2))^2+(r/4)^2}=\frac{\sqrt{13}}{2}r<2r.
    \end{equation}
    Therefore, two $H_{r/2}$ can lie completely within $H_r$. Now, if we define $X=\bigcup\limits_{i=1}^3 H^i_{r/2}$ where each $H^i_{r/2}$ $\forall$ $i$ is a hexagon of side $r/2$, and all three are mutually non-overlapping to each other, the furthest distance, in this case, is at least
    \vspace{-2mm}
     \begin{equation}
         l(X)=2\sqrt{(\sqrt{3}(r/2))^2+(r/4)^2}=\frac{\sqrt{13}}{2}r<2r.
     \end{equation}
     As shown in Fig. \ref{con_fig}, this configuration also lies within $H_r$. Note that, we cannot place four $H_{r/2}$ inside $H_r$, where all of them are mutually non-overlapping to each other. The reason for this is, that in this case, $l(X)=r+r+\frac{1}{2}r=\frac{5}{2}r > 2r$, which follows from Fig. \ref{con_fig}. Since we have $l(X) > 2r$, we claim that four $H_{r/2}$ cannot completely lie within $H_{r}$. 
\end{proof}

\begin{lemma}\label{lm2}
    If sensor $s_i$ with sensing range $SR_i$ is placed on a vertex of an equilateral triangle of side $r$, the entire triangle lies within $SR_i$.
\end{lemma}

\begin{proof}
    Let $\Delta ABC$ be an equilateral triangle of side $r$ and a sensor $s_i$ with sensing range $SR_i$ is placed on $A$. If we draw a circle of radius $r$ at $A$, the vertices $B$ and $C$ lie within its circumference. Similarly, $A, C$ and $A, B$ lie on the circles' circumference, whose center is at $B$ and $C$, respectively. Since, the circles center at $A, B$, and $C$ are convex, their intersecting region formed by $A, B$, and $C$ is also a convex zone \cite{cana}. Moreover, $\Delta ABC$ also lies on the convex zone formed by $A, B, $ and $C$. Therefore, $\Delta ABC$ lies in $SR_i$.
\end{proof}

Note that, for $s_i$, both $SR_i$ and $CR_i$ are circular disks of a certain radius. But we can not cover an Euclidean plane with circular tiles. It is well-studied that a hexagonal tile can cover the greatest area inside a circular disc \cite{k-cov}. As a result, we use the hexagonal tiles to cover the Euclidean plane.

\subsection{Sensor Deployment Strategy}\label{dply}
As both $SR_i$ and $CR_i$ can be approximated as hexagonal tiles, the sensors can cover the Euclidean plane in a way similar to these tiles. Moreover, here we propose a sensor deployment strategy that enables us to obtain $k$-coverage, but with fewer sensors. Notably, we strategically place the sensor inside a regular hexagon, and by employing this method for every hexagon, we can cover the entire Euclidean plane.
 In this context, we classify our strategy into four separate cases, as described below.
 \begin{figure*}[t!]
     \centering
     \includegraphics[width=.72\linewidth]{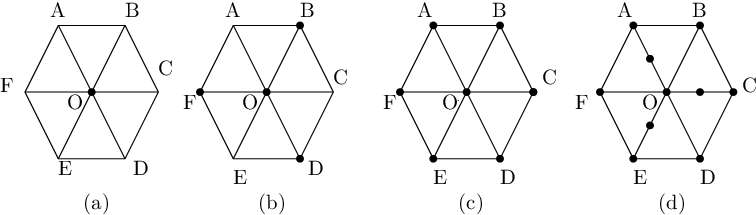}
     \caption{Sensor Deployment Strategy}
     \vspace{-4mm}
     \label{sdply}
 \end{figure*}
 \subsection*{A-I: For $k=1$ ($1$-Coverage)}
In this case, as shown in Fig.\ref{sdply}(a), we deploy the sensor at the center $O$ of the hexagon. Note that, as $O$ is the common vertex of all the six equilateral triangles contained in the hexagon, this deployment strategy is adequate for providing $1$-coverage to these triangles, which also follows from Lemma \ref{lm2}. Note that, $k$-coverage can easily be obtained by placing $k$ co-located sensors at $O$. 

Before discussing the higher coverage sensor deployment strategy for a hexagon, we demonstrate in the ensuing lemma that, to obtain $2$-coverage, where co-located sensors are prohibited, at least three additional sensors are required.

\begin{lemma}
    If placing a sensor in the center $O$ is prohibited, at least three sensors are required to cover a hexagon.
\end{lemma}
\begin{proof}
    A hexagon consists of six equilateral triangles and two consecutive triangles within the hexagon share a common edge. Moreover, from Lemma \ref{lm2}, it follows that a point on or within a triangle can cover the entire triangle. It is evident that, the entire triangle cannot be covered by a point lying outside of it. Therefore, if we deploy a sensor on the common edge of two triangles it can cover both of them. Moreover, as a hexagon contains six triangles, we can assert that at least three sensors are necessary for covering the hexagon.
\end{proof}

\subsection*{A-II: For $k=2$ ($2$-Coverage)}
As shown in Fig.\ref{sdply}(b), here also, there are six equilateral triangles within the hexagon. Consequently, to achieve $2$-coverage of all these triangles, we need to deploy the sensors in a way such that each triangle lies inside the coverage area of at least two sensors. From the figure, we observe that $O$ is the center, and $A, B, C, D, E, F$ are the vertices of the hexagon. As discussed previously, deploying a sensor at $O$ provides $1$-coverage for all the triangles. Therefore, if the sensors are deployed on the alternate vertices, i.e., at $B, D,$ and $F$; or at $A,C,$ and $E$, this, along with the sensor placed at $O$, provides $2$-coverage to all the equilateral triangles.

\subsection*{A-III: For $k=3$ ($3$-Coverage)}
Unlike the previous case, here we strategically deploy the sensors at all the vertices of the hexagon, i.e., $A,B,C,D,E,$ and $F$, and also at the center $O$. As a result of which, Fig.\ref{sdply}(c) demonstrates that all the triangles lie inside the coverage area of at least $3$ sensors, i.e., a $3$-coverage scenario.

Note that, up to the case of $k=3$-coverage, we deployed the sensor on the vertices and at the center of the hexagon. However, we can not place more than one sensor at the same place. Therefore, for the case of $k>3$-coverage scenario, we strategically place the sensors inside the regular hexagon, as discussed next.

\subsection*{A-IV: For $k>3$ ($k>3$-Coverage)}
In Fig.\ref{sdply}(d), we observe that $A, B, C, D, E,$ and $F$ are the vertices of the hexagon. Also, $OA$ and $OB$ are the sides of $\Delta OAB$, $OC$ and $OD$ are the sides of $\Delta OCD$, and $OE$, $OF$ are the sides of $\Delta OEF$, respectively. Accordingly, we rename $OA,OB,OC,OD,OE$, and $OF$ as $l_1, \cdots,l_6$.  Therefore, to get $k$-coverage $(k>3)$ from $(k-1)$-coverage, we deploy $3$ additional sensors strategically on $l_i, i=1,\cdots,6$. Specifically, for an even (odd) $k$, we deploy the sensors on those $l_i$, whose index is odd (even). For example, for $k=4$, we deploy three sensors on $l_1, l_3 $ and $l_5$, respectively, i.e., on $OA, OC$ and $OD$ which is shown in Fig.\ref{sdply} (d). Similarly, for $k=5$, we deploy three sensors on $l_2, l_4 $ and $l_6$, respectively, i.e., on $OB, OD$ and $OF$. Note that, we can deploy the sensor at any point on the line but doing so at the same location is not allowed.

\subsection{Required Number of Sensors for $k$-coverage}\label{no.of_sen}
In Fig.\ref{sdply}(a), we observed that  $1$-coverage can be achieved with a single sensor placed at the center of the hexagon. Note that, there are six triangles inside the hexagon and any two consecutive triangles share a common side. Consequently, it is beneficial for both triangles to achieve $k$-coverage if we place sensors on their common sides. Therefore, four sensors are sufficient for $2$-coverage  and eventually, for $3$-coverage, $7$ sensors are adequate. As a result, we can state, that if $n_1$ sensors are required for $k$-coverage for $k >3$, then $n_1+3$ sensors are adequate for $(k+1)$-coverage. Therefore, we can express this as an arithmetic progression, whose first term is $1$ and the common difference is $3$. Hence, for the general $k$-coverage, the required number of sensors are
\vspace{-2mm}
\begin{equation}\label{sno}
    n_1=1+3(k-1)=3k-2.
\end{equation}

\subsection{Sensor Density per Unit Area }  \label{sdensity}
Concerning our proposed deployment strategy, here we calculate the respective sensor density per unit area. From \eqref{area}, we have the area of the hexagon of side $r$ to be $\frac{3\sqrt{3}}{2}r^2$. Furthermore, from \eqref{sno}, we claim that, $3k-2$ sensors are sufficient to support $k$-coverage within the hexagon. Hence, if $s_d$ is the sensor density per unit area, we obtain 
\vspace{-2mm}
\begin{equation}\label{den}
    s_d=\dfrac{3k-2}{\frac{3\sqrt{3}}{2}r^2}=\dfrac{2(3k-2)}{3\sqrt{3}r^2}.
\end{equation}
For example, when $k=1$, we get $s_d=\frac{2}{3\sqrt{3}r^2}$.

On the other hand, the authors in \cite {k-cov} have discussed another sensor deployment strategy for an Euclidean plane. They used regular hexagons of side $r/2$ to cover an Euclidean plane. As a result, for $k$-coverage, they randomly placed $k$ sensors, each having sensing range $r$, inside the hexagon of side $r/2$. In their proposed strategy, the respective sensor density per unit area is
\begin{equation}  \label{den_othr}
    s_d^{\rm ex}=\frac{k}{\frac{3\sqrt{3}}{2}(r/2)^2}=\frac{8k}{3\sqrt{3}r^2}.
\end{equation}


Accordingly, by using \eqref{den} and \eqref{den_othr}, we quantify the gain of our proposed strategy as
\vspace{-2mm}
\begin{equation}
    s_d^{\rm ex}-s_d=\frac{8k}{3\sqrt{3}r^2}-\frac{2(3k-2)}{3\sqrt{3}r^2}=\frac{2k+4}{3\sqrt{3}r^2}.
\end{equation}
Thus, we observe that for any positive $k$, $s_d^{\rm ex}-s_d>0$, i.e., our strategy outperforms the existing benchmark. Another interesting insight, which can be obtained is
\vspace{-2mm}
\begin{equation}
    \lim_{k \rightarrow \infty} \frac{s_d}{s_d^{\rm ex}}=\frac{3}{4},
\end{equation}
which demonstrates the advantage of our strategy as $k \rightarrow \infty$.

\subsection{Required Number of Sensor for a Plane}  \label{rqd_sens}
In Section \ref{dply}, we discussed our sensor deployment strategy for a single hexagon with side $r$. As shown in Fig.\ref{sys}, the entire plane can be tiled using the regular hexagon of side $r$. So, we looked into the required number of sensors for $k$-coverage of the entire plane. 


To achieve this, we look into a model with one central hexagon and six more hexagons around it. Each of these second `layer' of hexagons shares an edge with the six edges of the central one. By continuing with this procedure, we obtain the third, fourth, and consecutive layers of hexagons, with respect to the central one. Let $l$ denote the number of layers that cover the plane. We name this `the solar model', as is shown in Fig. \ref{sys}, where  $O$ denotes the center of the central hexagon.  Accordingly, we obtain a closed-form analytical expression for the required number of sensors to support $k$-coverage upto $l$ layers. In this context, we classify our strategy into four separate cases, as described below.

 \subsection*{D-I: For $k=1$ ($1$-Coverage)}
 From previous discussions, we observe that in $1$-coverage a single sensor lies at the centre of the hexagon. Thus,  the required number of sensors, in this case, is equal to the total number of hexagons in the solar model. Therefore, from Fig.\ref{sys}, we observe that a single hexagon lies in the first layer, $6$ hexagons lie in the second layer, and in general, $6(l-1)$ hexagons lie in the $l$-th layer, where $l\geq 2$. Specifically, the number of hexagons lie in the $l$-th layer for $l \geq 2$, follows an arithmetic progression whose first term as well as common difference is $6$. Hence the total number of hexagons that lie within the solar model is 
\begin{align}\label{layer}
    1+6+12+ \cdots +6(l-1) =1+3l(l-1).  
\end{align}
Therefore, in this case, $n(l,k=1)$, the required number of sensors to get $1$-coverage of $l$ layers is given by
\vspace{-2mm}
\begin{equation}  \label{nk1}
    n(l,k=1)=1+3l(l-1)).
\end{equation}

\subsection*{D-II: For $k=2$ ($2$-Coverage)}\label{2-cov-gen}

To achieve $2$-coverage, we consider that  $n(l,k=1)$ sensors are already deployed to support $1$-coverage. Section III-A shows that to support $2$-coverage, we need three more sensors in the central hexagon, placed at the alternate vertices, apart from the one at the center. Similarly, for the second layer of hexagons, we need $9$ sensors and for the third layer, $15$ sensors are required. Moving forward, we need $3+6(l-1)$ sensors for the $l$-th layer. Therefore,  we obtain the total number of sensors required as
\begin{align}  \label{nk2}
    n(l,k=2)&=n(l,k=1)+(3+9+\cdots+(3+6(l-1)) \nonumber \\
    &=n(l,k=1)+\frac{l}{2}\left(2\times3+(l-1)6\right) \nonumber \\
    &\overset{(a)}{=}1+3l(l-1))+3l^2=6l^2-3l+1,
\end{align}
where $(a)$ follows from \eqref{nk1}.

\subsection*{D-III: For $k=3$ ($3$-Coverage)}
Similar to the previous discussion, we observed that having $3$-coverage implies placing sensors on all the vertices and at the center of the hexagon. Alternatively, for $3$-coverage scenario up to $l$ layers, as discussed earlier, we consider that $n(l,k=2)$ sensors are required to support $2$-coverage. Note that if we place the sensors on the remaining sensor-free vertices, the $l$ layers achieve the $3$-coverage. Here, the number of remaining sensor-free vertices is
\vspace{-2mm}
\begin{equation}
    \frac{l}{2}\left(2\times3+(l-1)6\right)=3l^2.
\end{equation}
This can be proved in a way similar to the process as discussed for the case of $2$-coverage. Thus, the required number of sensors, in this case, is
\vspace{-2mm}
\begin{equation}  \label{nk3}
    n(l,k=3)=n(l,k=2)+3l^2=9l^2-3l+1.
\end{equation}

\subsection*{D-IV: For $k>3$ ($k>3$-Coverage)}
Up to $3$-coverage, sensors are strategically deployed on the the vertices, and at the center of the hexagons. But, after $3$-coverage, they are deployed inside the hexagons, and to increase one more coverage, three more sensors are needed for every single hexagon. Hence, for each $k>3$, to get $k$-coverage, $3\times(1+3l(l-1))$ more sensors are needed as compared to $n(l,k-1)$. Therefore, the total number of sensors required is
\begin{align}  \label{nkabv3}
    n(l,k)&=n(l,k=3)+(k-3)(3\times(1+3l(l-1))) \nonumber \\
    &=9(k-2)l^2-3(3k-8)l+3k-8
\end{align}
By combining \eqref{nk1}, \eqref{nk2}, \eqref{nk3}, and \eqref{nkabv3}, we obtain an unified expression for $n(l,k)$ as
\vspace{-2mm}
\begin{equation}
    \!\!\!\!n(l,k)=\begin{cases}
        1+3l(l-1) & k=1, \\
        6l^2-3l+1 & k=2, \\
        9(k-2)l^2-3(3k-8)l+3k-8 & k \geq 3.
    \end{cases}
\end{equation}
On the other hand, by using Lemma \ref{lema1} and the existing sensor deployment strategy as discussed in \cite{k-cov}, the associated total number of sensors required to get $k$-coverage for $l$ layers is
\vspace{-2mm}
\begin{equation}
    \!\!\!\!n^{\rm ex}(l,k)=\begin{cases}
        1+3l(l-1) & k=1, \\
        k(15l^2-27l+18) & k\geq 2. 
    \end{cases}
    \end{equation}
The expression for $k\geq 2$ is obtained based on the fact that $15l^2-27l+18$ is the total number of $H_{r/2}$ completely lies inside the solar model of $l$ layer where each $H_{r/2}$ contains $k$ sensors. 
It is interesting to observe, that irrespective of $k$ and $l$, we always have $n^{\rm ex}(l,k)-n(l,k)\geq 0$, which demonstrates the benefit of our proposed strategy. Another interesting insight, which can be obtained is
\begin{equation}
    \lim_{k \rightarrow \infty } \lim_{l \rightarrow \infty } \frac{n(l,k)}{n^{\rm ex}(l,k)}=\frac{3}{5},
\end{equation}
which demonstrates the advantage of our strategy as $k,l \rightarrow \infty$.

\section{Numerical Results}
In this section, we evaluate the proposed sensor deployment strategy and also, compare its performance with that of the existing benchmark scheme in \cite{k-cov}. As stated earlier, the work in \cite{k-cov} follows a computation geometry-based approach (CGA), where the authors suggest to deply the sensors in a certain manner in order to attain $k$-coverage.
\begin{figure}[t!]
    \centering
    \includegraphics[width=0.76\linewidth]{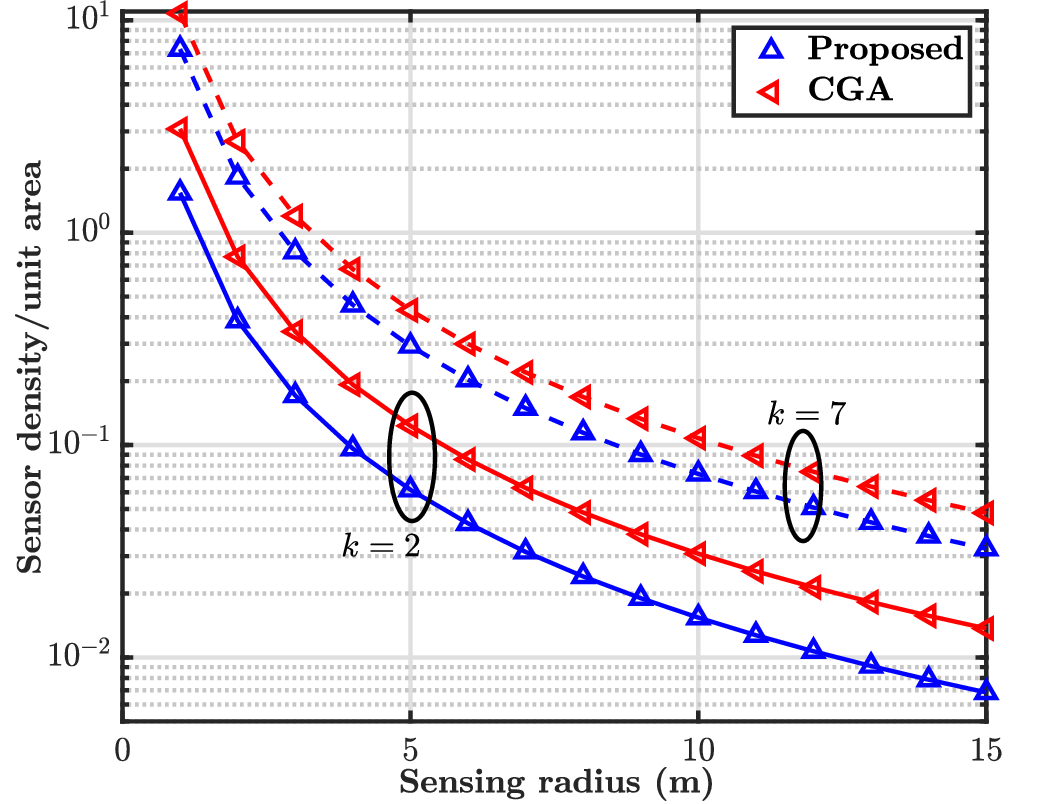}
    \vspace{-2mm}
    \caption{Impact of sensing radius on sensor density.}
    \label{den_r}
    \vspace{-3.2mm}
\end{figure}

Fig. \ref{den_r} depicts the impact of the sensing radius on the deployed sensor density. Here we demonstrate the effect of both the proposed strategy and CGA specifically to attain $k$-coverage for $k=2,7$. We observe that while the sensor density decreases monotonically with the sensing radius irrespective of the approach being considered, the proposed strategy results in significantly lower sensor density, when compared against CGA. The reason for this is attributed to the intelligent placing of sensors on the vertices and edges of the hexagons.

\begin{figure}[t!]
    \centering
    \includegraphics[width=0.76\linewidth]{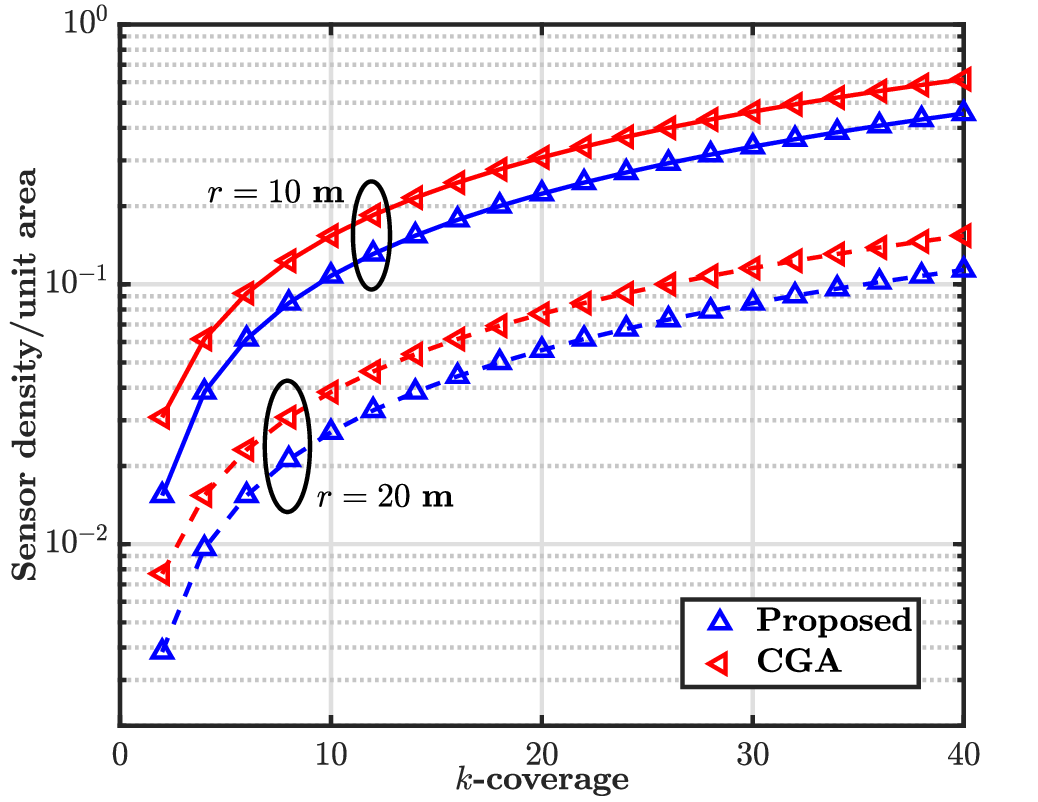}
    \vspace{-2mm}
    \caption{Impact of $k$-coverage on sensor density.}
    \label{dev_k}
    \vspace{-3.2mm}
\end{figure}

Fig. \ref{dev_k} illustrates the variation of sensor density with the $k$-coverage, for both the cases of $r=10$ m and $r=20$ m, respectively. While we observe that in both the cases, the sensor density increases with the value of $k$ in $k$-coverage, the reason is intuitive. A higher value of $k$ implies that, at any point of time, any particular area is being covered by $k$ sensors. However, at the same time, we observe that with identical system parameters, our proposed approach results to a lower sensor density as compared to CGA.

\begin{figure}[t!]
    \centering
    \includegraphics[width=0.76\linewidth]{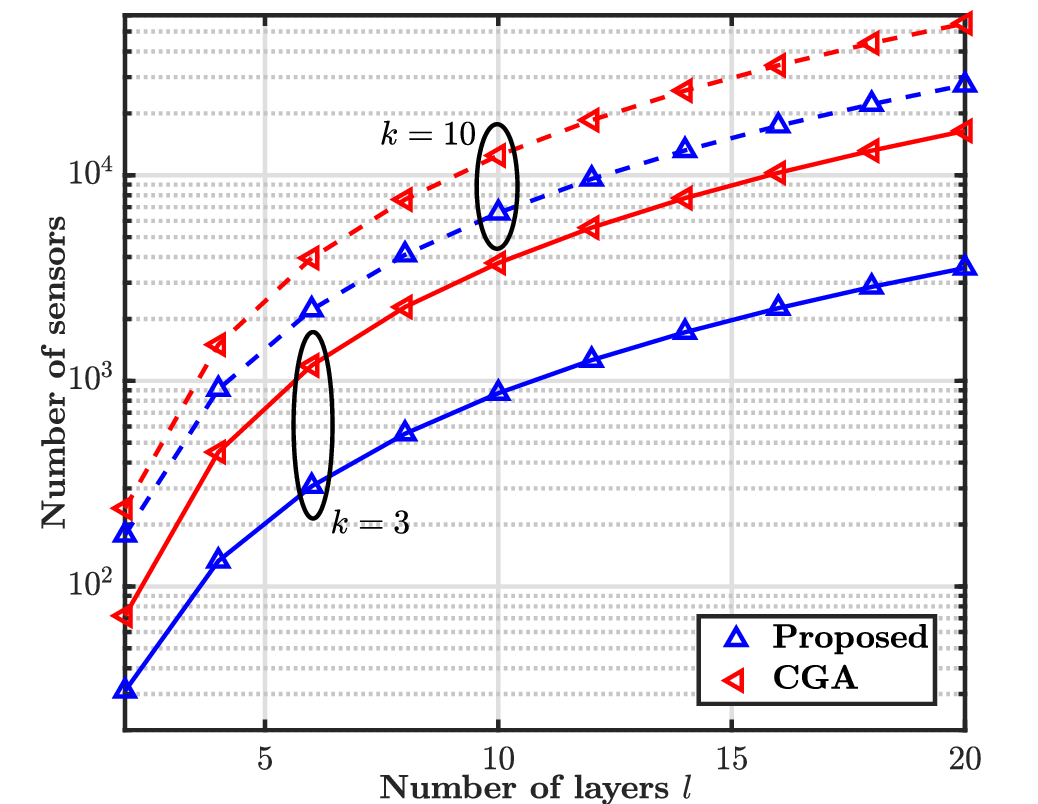}
    \vspace{-2mm}
    \caption{Impact of layers on required number of sensors.}
    \label{sen_lay}
    \vspace{-3.2mm}
\end{figure}

\begin{figure}[t!]
    \centering
    \includegraphics[width=0.76\linewidth]{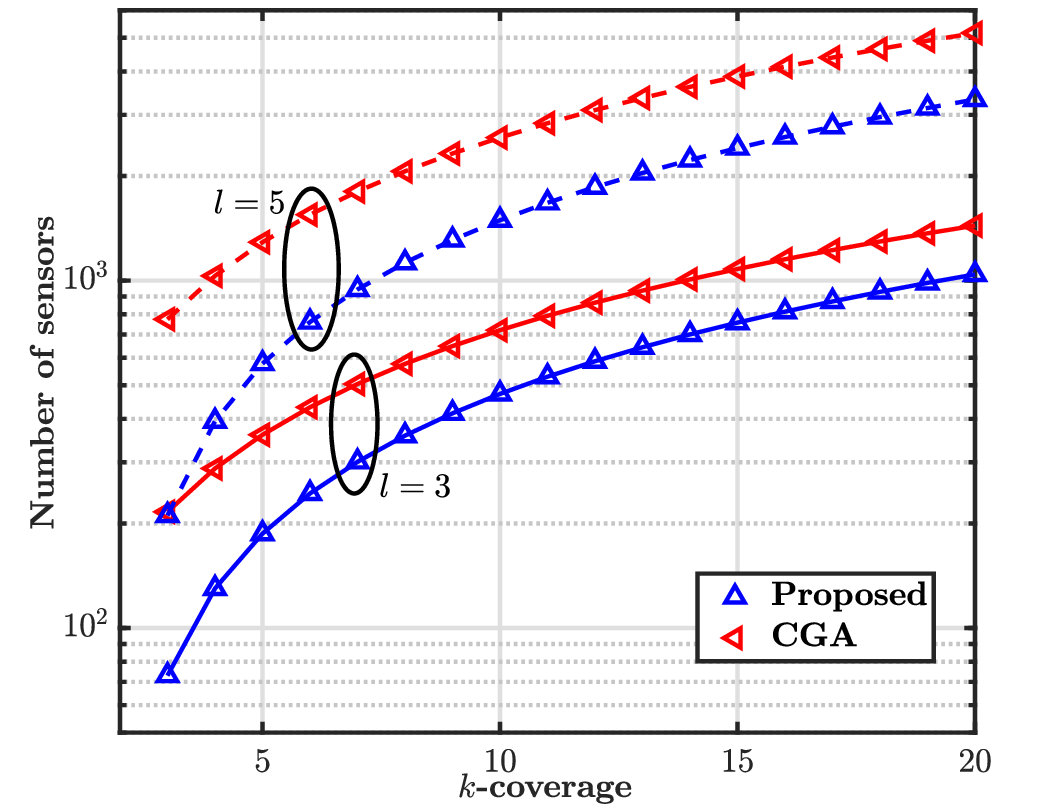}
    \vspace{-2mm}
    \caption{Impact of $k$-coverage on required number of sensors.}
    \label{nsen_k}
    \vspace{-3.2mm}
\end{figure}

For a given number of layers $l$ and $k$-coverage, Section \ref{rqd_sens} obtains a closed-form expression for the required number of sensors $n(l,k)$. Accordingly, Fig. \ref{sen_lay} investigates the impact of $l$ on $n(l,k)$ for $k=3$ and $k=10$. Similarly, Fig. \ref{nsen_k} illustrates the impact of $k$ on $n(l,k)$ for $l=3$ and $l=5$, respectively. In both these figures, we observe an increasing trend irrespective of the scheme employed, which is intuitive. Moreover, they also demonstrate the enhanced performance of the proposed sensor deployment scheme against CGA. As explained earlier, here also, this performance gain is attributed to the intelligent placement of sensors to obtain the desired coverage with a certain number of layers.

\begin{figure}[t!]
    \centering
    \includegraphics[width=0.76\linewidth]{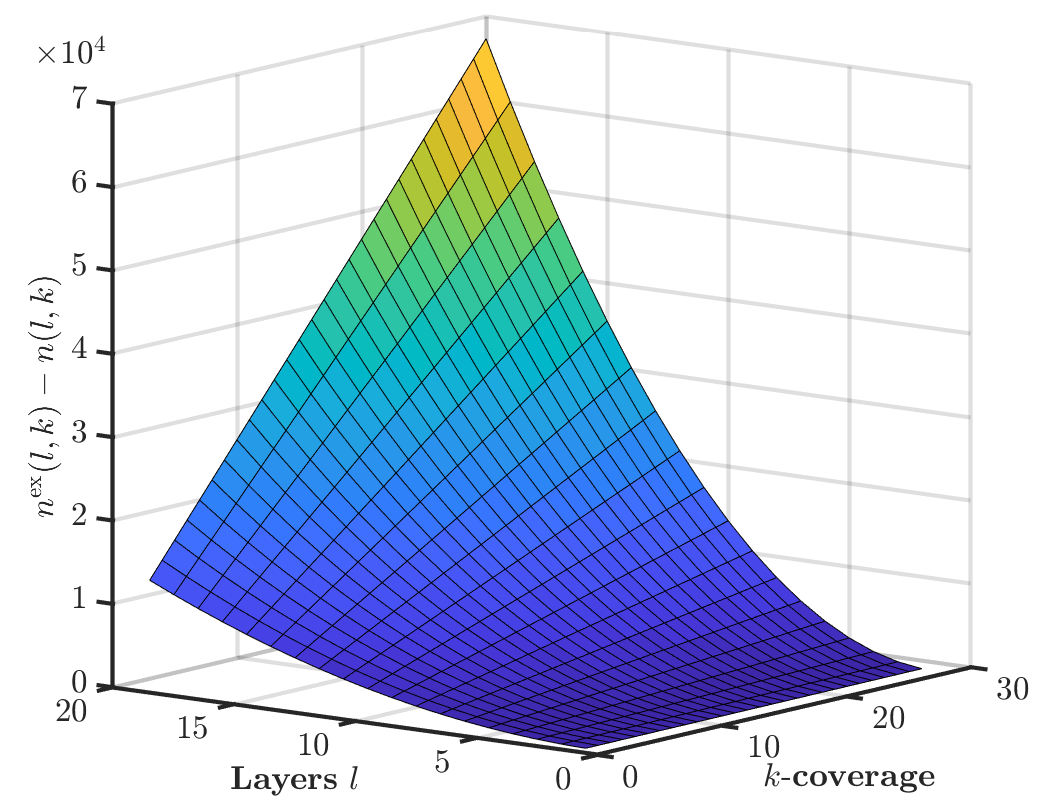}
    \vspace{-2mm}
    \caption{Joint effect of $k$-coverage and number of layers on required number of sensors.}
    \label{nsen_kl}
    \vspace{-3.2mm}
\end{figure}

Finally, Fig. \ref{nsen_kl} demonstrates the joint effect of coverage and number of layers on the number of sensors required. Specifically, by varying both $l$ and $k$, we calculate both $n(l,k)$ and $n^{\rm ex}(l,k)$ according to our proposed strategy and CGA, respectively. Henceforth, we plot $n^{\rm ex}(l,k)-n(l,k)$ by jointly varying both the parameters. We observe from the figure that this quantity is always positive in nature, i.e., our deployment strategy requires lesser sensors compared to CGA with identical scenarios. Moreover, the variation with $l$ for a given $k$ is relatively sharp compared to the variation with $k$ for a given $l$. This is because both $n(l,k)$ and $n^{\rm ex}(l,k)$ are linear and quadratic with respect to $k$ and $l$, respectively.

\section{Conclusion}
In this work, we investigated a graph theoretic sensor deployment approach in WSNs. Specifically, we modeled the sensor deployment problem as a $k$-coverage problem to propose a respective solution. Accordingly, we obtained analytical closed-form expressions of two performance metrics, namely `sensor density per unit area' and `required number of sensors for a plane' in terms of the sensing radius $r$, coverage $k$, and level $l$, respectively. Also, we evaluated these quantities for the existing benchmark schemes to quantify the advantages of our proposed strategy. We demonstrated that the required number of sensors is linear and quadratic with respect to $k$ and $l$, respectively. Finally, the numerical results validate the advantages of our proposed strategy. Based on the framework presented in this paper, an immediate extension of this work is to investigate the scenario, where tiling does not always necessarily happen by regular hexagons.

\bibliographystyle{IEEEtran}
\bibliography{ref}

\end{document}